\documentclass{ws-mpla}
\usepackage{feynmf}
\input slashed.sty
\usepackage{slashed}
\begin{document}


\catchline{}{}{}{}{}

\title{Quenching of the Deuteron in Flight\footnote{Supported in part by the Forschungszentrum J\"ulich (COSY)}}

\author{\footnotesize M. Dillig}

\address{Institute for Theoretical Physics III \footnote{preprint FAU-TP3-06/Nr. 09}, University of Erlangen-N\"urnberg,\\
Staudtstr. 7, D-91058 Erlangen, Germany\\
mdillig@theorie3.physik.uni-erlangen.de}

\author{C. Rothleitner}

\address{Max Planck Research Group, Institute of Optics, Information and Photonics, \newline University of Erlangen-N\"urnberg, \\
G\"unther-Scharowsky-Str. 1/Bau 24, D-91058 Erlangen, Germany\\
crothleitner@optik.uni-erlangen.de}

\maketitle


\begin{abstract}
We investigate the Lorentz contraction of a deuteron in flight. Our starting point is the Blankenbecler-Sugar projection of the Bethe-Salpeter equation to a 3-dimensional quasi potential equation and apply it for the deuteron bound in an harmonic oscillator potential (for an analytical result) or by the Bonn NN potential for a more realistic estimate. We find substantial quenching with increasing external momentum and a significant modification of the high momentum spectrum of the deuteron.

\end{abstract}

\ccode{PACS: 03.65.Pm, 11.30.Cp, 12.39.Ki}

\section{Introduction}
One of the main goals of physics with modern accelerators is the detailed investigation of the
response of the nuclei\cite{1Machner} to large momentum transfers of around or above $1\; GeV\slash
c$
. In particular, here the lightest nucleus, i.e. the deuteron, is a preferred object due to its well known nuclear structure\cite{2Garcon}. In recent years many investigations with the deuteron as a target have been performed at large momentum transfers with proton\cite{3Buscher} and electron beams\cite{4Osipenko}; further studies with deuteron beams have been\cite{5Terrien} or will be performed\cite{6Yonehara} in the near future. Here the deuteron as a spin 1 object is very special, as it allows to measure in addition to the total and differential cross section and the analyzing power also the tensor polarization of deuteron beams\cite{7Arvieux}. It is hoped -- and that explains the genuine interest -- that high momentum transfer processes reveal details of the short range reaction mechanism and of the deuteron itself, which are expected at momentum transfers up to the low $GeV\slash c$ region to be dominated by meson exchange currents\cite{8Chemtob} and the excitation of baryon resonances\cite{9Burkert}.\\

In most calculations the deuteron is conventionally described as an object in rest with standard s and d wave components. However it is clear that with increasing momenta or momentum transfers on the deuteron this picture has to be modified: in coordinate space a moving deuteron is quenched along its axis of motion. As a consequence the deformation of the deuteron is increasing with increasing momenta which leads to a significant modification and enhancement of the deuteron spectrum at high momentum components. This, as expected, will modify substantially predictions of observables, which depend sensitively on large momenta components of the deuteron.\\

In standard calculations the quenching of the deuteron is frequently not included, instead one uses approximate descriptions, such as to go to the Breit frame for the deuteron form factor\cite{10Feynman}, in order to hopefully minimize the error due to the approximation.\\

In this note we investigate the quenching of a moving deuteron in a simple model. In chapter 2 we start from the manifest covariant Bethe-Salpeter equation\cite{11Salpeter} and solve, after projecting onto 3 dimensions, a still covariant quasi potential equation\cite{12Erkelenz} for an harmonic oscillator potential of the deuteron with the goal to obtain analytical solutions. These solutions are then derived explicitly within a few approximations for the d-state of the deuteron, and presented in chapter 3, together with characteristic results for the deuteron wave function in momentum space; subsequently we apply the same prescription to a realistic deuteron wave function from the Bonn potential. Finally we close in chapter 4 with a summary and some remarks to improve of the approach.

\section{Basic equations}

In this chapter we derive for a given kinematics the quasi potential equation for the deuteron from a Blankenbecler-Sugar projection of the Bethe-Salpeter equation (BSE).

\subsection{Bethe-Salpeter equation for the deuteron}

For the deuteron with the total momentum $P$ and the relative momentum $q$ between the two nucleons the BSE is given as

\begin{equation}
\psi(P,q)=\frac{(\frac{\slashed{P}}{2}+\slashed{q}+M)(\frac{\slashed{P}}{2}-\slashed{q}+M)}
{((\frac{P}{2}-q)^2-M^2-\text{i}\eta)((\frac{P}{2}+q)^2-M^2-\text{i}\eta)}\\
\int{K(P,q,k)\psi(P,k)dk}
\label{glg1}
\end{equation}
where $K(P,q,k)$ denotes the kernel of the BSE and $M$ the nucleon mass. The total 3-momentum $\underline{P}$ is specified along the z-axis, i.e. $\underline{P}=(0,0,P_z)$, while the 4-momentum $P$ is given by $P=(E_d,0,0,P_z)$ with the total energy of the deuteron given as $E_d= \sqrt{M^ 2_d +P^ 2_z}$ ($M_d$ is the deuteron mass). To simplify the complicated spinor structure of $\psi (P,q)$, which is a 16 component spinor, we neglect in $M_d=2M+\varepsilon_d$ and $2M\gg \varepsilon_d$ the binding energy of the deuteron $\varepsilon_d\cong -2.2\: MeV$, and put the two nucleons on the mass shell; then $(\tfrac{\slashed{P}}{2}\pm \slashed{q}+M)\cong 2M$ (this corresponds in keeping only the large components in each of the Dirac spinors of the two nucleons). With this approximation the BSE is simplified drastically to
\begin{align}
\psi(P,q)=&\frac{4M^2}{((\frac{P}{2}+q)^2-M^2-\text{i}\eta) ((\frac{P}{2}-q)^2-M^2-\text{i}\eta)}\nonumber\\
&\times\int K(P,q,k)\psi(P,k)dk\, .
\label{glg2}
\end{align}

\subsection{The Blankenbecler-Sugar (BS) projection}

As the BSE can -- except for separable potentials and a few academic cases -- only be solved numerically, we simplify the problem by a reduction of the BSE to a 3-dimensional, but still covariant quasi-potential equation (QPE) by restricting the nonstatic effects in the BSE. Unfortunately the transition from 4 to 3 dimensions is not unique: there exists an infinite number of QPE\cite{13Yaes}. Thus we have to invoke -- guided by the underlying physics -- plausible arguments for the reduction of the BSE. As the two nucleons in the deuteron are expected to exhibit a symmetrical dependence on the fourth component of internal momentum $q$, we put them, following the arguments of Blankenbecler and Sugar\cite{14Blankenbecler}, in their Greens function
\begin{equation}
G(P,q) =\frac{4M^2}{\big{(}(\frac{P}{2}+q)^2-M^2-\text{i}\eta\big{)}\big{(}(\frac{P}{2}-q)^2-M^2-\text{i}\eta\big{)}}\\
\label{glg3}
\end{equation}
symmetrically off their mass shell
\begin{align}
\Rightarrow &
\text{i}\pi\frac{4M^2}{\big{(}(\frac{P}{2}+q)^2-M^2-\text{i}\eta\big{)} +\big{(}(\frac{P}{2}-q)^2-M^2-\text{i}\eta\big{)}}\nonumber\\
&\times
\delta\left(\frac{1}{2}\big{(}(\frac{P}{2}+q)^2-M^2\big{)}- \big{(}(\frac{P}{2}-q)^2-M^2\big{)}\right)
\label{glg4}
\end{align}
where the $\delta$-function can be solved for the $q_0$-component to give
\begin{equation}
\delta(Pq)= \frac{1}{E_d}\delta(q_0-\frac{P_z}{E_d}q_z).
\label{glg5}
\end{equation}
As a consequence the Greens function is now reduced to
\begin{equation}
G_{GP}(P,q)=2\text{i}\pi\frac{M}{E_d}\frac{\delta(q_0-\tfrac{P_z}{E_d}q_z)}{
\left(\frac{M_d^2}{4}-(\frac{M_d}{E_d})^2q_z^2-\underline{q}_\perp-M^2\right)}
\label{glg6}
\end{equation}
which depends explicitly on the external momentum $P_z$ of the deuteron. We verify that in the limit of a deuteron in rest, i.e. $E_d\rightarrow M_d=2M+\varepsilon_d$ the denominator $D$ in the Greens function recovers the standard non relativistic structure
\begin{align}
D&= \left(\frac{4M^2+4M\varepsilon_d}{4}-q_z^2-\underline{q}_\perp^2-M^2\right)\nonumber\\
&= M(\varepsilon_d-\frac{\underline{q}^2}{M}).
\label{glg7}
\end{align}
\newline
In this picture the full deuteron Greens function in the QP limit is then given dropping terms quadratic in $\varepsilon_d$, as
\begin{align}
G(P,\underline{q})&=\text{i}\pi\frac{4M^2}{2E_d}\delta(q_0-\frac{P_z}{E_d}q_z)
\frac{1}{(M\varepsilon_d-(\frac{M_d}{E_d})^2q_z^2-\underline{q}_\perp^2)}\nonumber\\
&\cong \text{i}\pi
M(\frac{M_d}{E_d})\frac{\delta(q_0-\frac{P_z}{E_d}q_z)}{(M\varepsilon_d-(\frac{M_d}{E_d})^2q_z^2-\underline{q}_\perp^2)}.
\label{glg8}
\end{align}
With these steps the structure of the QPE is, for a given interaction kernel $K(P,q,k)$, completely specified (the final expression for the QPE is given in the following chapter 3).

\section{Momentum spectrum of a deuteron in flight}
In this chapter we derive the structure of the momentum dependent harmonic kernel and present the solution of the corresponding QPE both for the central and the tensor piece of the deuteron wave function in momentum space.

\subsection{Harmonic kernel in momentum space}
As already mentioned we use for the potential of the deuteron an harmonic kernel, in order to end up with analytical solutions of the QPE. Thus we start from the four-dimensional form
\begin{equation}
K(x)\equiv V(x)=\lambda_{HO}x^ 2
\end{equation}
or its Fourier transform
\begin{align}
V(q-k)&=\lambda_{HO}\int x^2e^{\text{i}(q-k)x}dx\nonumber\\
&=-(2\pi)^4\lambda_{HO}\frac{d^2}{dq^2}\delta (q-k).
\label{glg9}
\end{align}
To reduce the 4 dimensional $\delta$-function to a 3-dimensional form we start from the representation\cite{15Singh}
\begin{equation}
\frac{d^2}{dq^2}\delta(q-k)=\lim_{m\rightarrow0}\frac{\partial^3}{\partial m^3}\frac{1}{(q-k)^2-m^2-\text{i}\eta}\\
\label{glg11}
\end{equation}
or, upon performing the BS-projection
\begin{equation}
\frac{d^2}{dq^2}\delta(q-k)=-\lim_{m\rightarrow 0}\frac{\partial^3}{\partial m^3}\frac{1}{(\frac{M_d}{E_d})^2(q_z-k_z)^2+(\underline{q}_\perp-\underline{k}_\perp)^2+m^2}
\label{glg12}
\end{equation}
so that the harmonic interaction kernel $V(q-k)$ reads with the notation
\begin{equation}
\quad \frac{M_d}{E_d} = \frac{M_d}{\sqrt {(M_d^2 + P_z^2)}} = \lambda
\label{glg13}
\end{equation}
\begin{equation}
V(\underline{q}-\underline{k})=(2 \pi)^4 \lambda_{HO}\frac{E_d}{M_d}\left(\frac{1}{\lambda^2}\frac{d^2}{dq_z^2}+\frac{d^2}{d\underline{q}_\perp^2}\right)\delta(\underline{q}-\underline{k}).
\label{glg14}
\end{equation}

\subsection{Deuteron wave function}
For a deuteron in rest the total wave function of the deuteron is given schematically as a combination of s and d wave pieces
\begin{equation}
\psi(E_d,\underline{q})=c_s\psi_s(q)+c_dS_{12}(\hat{\underline{q}})\psi_d(q)).
\label{glg15}
\end{equation}

\noindent Thereby $c_s$ and $c_d$ define the strength of the s and d-wave, which depend on the underlying nucleon-nucleon potential; with the tensor operator \newline  $S_{12}(\underline{q})=3\underline{\sigma_1}\underline{q}\underline{\sigma_2}\underline{q}-\underline{\sigma_1}\underline{\sigma_2}$, this yields in a coupled form
\begin{equation}
\psi(E_d,\underline{q})=\left( c_sR_s(q)[Y_0(\hat{\underline{q}})[\tfrac{1}{2}\tfrac{1}{2}]^1]^ {1m}+c_dR_d(q)[Y_2(\hat{\underline{q}})[\tfrac{1}{2}\tfrac{1}{2}]^1]^{1m}\right)[\tfrac{1}{2}\tfrac{1}{2}]^{0,0}
\end{equation}
with radial functions $R_{s,d}(q)$ (the coupled form $[\tfrac{1}{2}\tfrac{1}{2}]^{0,0}$ denotes the isospin content of the deuteron). In the following we discuss the solution for the s and the d-wave separately.

\subsection{Solution for the s-state}
With the steps given above, we obtain from Eqs. (8,15) and with the additional notation
\begin{equation}
\Lambda^2=\text{i}(2\pi)^5\lambda_{HO}
\label{glg16}
\end{equation}
the final equation for the s-wave in momentum space
\begin{equation}
\left(\frac{\Lambda^2}{\lambda^2}\frac{d^2}{dq_z^2}+\Lambda^2\frac{d^2}{d\underline{q}_\perp^2}-M\varepsilon_d-(\frac{M_d}{E_d})^2q_z^2-q_\perp^2\right)\psi(P,\underline{q})=0.
\label{glg17}
\end{equation}
Note that the notion s-wave is somewhat misleading, as the motion of the deuteron along the z-axis breaks the rotational symmetry of the standard s-wave component and introduces angular momenta $l\neq 0$.\\
\newline
\indent The solution of Eq. (18) is easily obtained, following standard text books. Going over to cartesian coordinates with
\begin{equation}
\left((\frac{\Lambda^2}{\lambda^2}\frac{d^2}{dq_z^2}+\Lambda^2\frac{d^2}{dq_x^2}+\Lambda^2\frac{d^2}{dq_y^2})-M\varepsilon_d \lambda^2 q^2_z-q^2_x-q^2_y)\right)\psi(P,q)=0\\
\label{glg18}
\end{equation}\\
we obtain with the notation
\begin{gather}\label{grp}
\lambda^2_z=\frac{\lambda^4}{\Lambda^2}, \quad
\lambda^2_x=\frac{1}{\Lambda^2}=\lambda_y^2\\
k_z^2=\frac{\lambda^2M}{\Lambda^2}\varepsilon_z,\quad k^2_{x,y}=\frac{M}{\Lambda^2}\varepsilon_{x,y}\\
\varepsilon_d=\varepsilon_x+\varepsilon_y+\varepsilon_z
\label{glg19}
\end{gather}
the final equations
\begin{align}
\left(\frac{d^2}{dq_z^2}+k^2_z-\lambda^2_zq^2_z\right)\phi(q_z)&=0\\
\left(\frac{d^2}{dq_i^2}+k^2_i-\lambda^2_iq^2_i\right)\phi(q_i)&=0
\label{glg20}
\end{align}
(where $i=x,y$). Then the solution for the radial part is given as
\begin{equation}
\phi(q_i)=N_ie^{\tfrac{1}{2}\lambda_iq^2_i}
\label{glg21}
\end{equation}
($N_i$ are normalization constants).
\newline

The modification of the momentum spectrum along the z-axis is easily seen in comparison with a deuteron in rest: comparing with the solution of the spherical harmonic oscillator
\begin{equation}
\phi\sim e^{\frac{a^2q^2}{2}}
\label{glg22}
\end{equation}
where $a$ is the common oscillator parameter, we find from
\begin{align}
\lambda_x&=\lambda_y=\frac{1}{\Lambda}=a^2\\
\lambda_z&=\frac{\lambda^2}{\Lambda}=(\frac{M_d}{E_d})^2a^2\: < \:\lambda_{x,y} \quad\text{for}\: P_z>0
\label{glg23}
\end{align}
immediately an enhancement of the z-components of a deuteron in flight or, correspondingly, a quenching of the z-component of the deuteron in coordinate space.

\subsection{Solution for the d-state}
The d-state component of a deuteron in flight is more complex than the corresponding s-state. A
rigorous solution, even for an harmonic kernel, can be obtained only numerically, as the tensor
operator couples longitudinal and perpendicular momentum components. We suppress this coupling by
the assumption that in
\begin{equation}
S_{12}(\underline{q})=\sqrt{24\pi}(q_z^2+\underline{q}^2_\perp)[Y_2(q_z,\hat{\underline{q}}_\perp)[\underline{\sigma}_1,\underline{\sigma}_2]^2]^{0,0}
\end{equation}
for $\mid\underline{q}_{\perp}\mid\ll q_z$ -- the case we are most interested in -- the unit vector for the total momentum is directed along the z-axis (with $Y_{20}=\sqrt{\tfrac{3}{4\pi}}$ and $ Y_{22}=\sqrt{\tfrac{15}{16\pi}}$) then only the terms $q^2_z$ and $\underline{q}^2_{\perp}$ without an explicit dependence on the angular momentum enter in the decoupled radial equations for $q_z$ and $\underline{q}_{\perp}$. Consequently we end up, after separating into cartesian coordinates, with the radial equations
\begin{equation}
(q^2_i\frac{d^2}{dq^2_i}+k^2_i-\lambda^2_i q^2_i)\phi(q_i)=0
\end{equation}
for the x,y and z components (following the notation for the s-state equation in 3.3). The
solutions of these differential equations are readily obtained: the general equation
\begin{equation}
x^2f''(x)+(b^2-a^2x^2)f(x)=0
\label{glg24}
\end{equation}
is reduced with suitable substitutions
\begin{equation}
f(x)=\sqrt{x}y(x),\quad u=\sqrt{-a^2}x
\label{glg25}
\end{equation}
to the differential equation for Bessel functions
\begin{equation}
u^2y''(u)+uy'(u)+(u^2-(\tfrac{1}{4}-b^2))y(u)=0
\label{glg26}
\end{equation}
with the corresponding regular solution
\begin{equation}
f(x)=\sqrt{x}J_{(\tfrac{1}{4}-b^2)}\left(\sqrt{-a^2}x\right).
\label{glg27}
\end{equation}
($J$ is a Bessel function of first kind.)

\subsection{Results and discussion}
In this section we present typical results from the formalism derived in the proceeding sections. Thereby we focus on the momentum spectrum of the s and d-state, as here the influence of the quenching of a deuteron in flight is seen most clearly. To model our static deuteron we use an oscillator parameter $a=0.7fm$ as a typical scale (of course we are aware that the representation of deuteron wave function as an harmonic oscillator with a single size parameter is an inadequate representation for a realistic comparison). Typical results are given in Fig. 1(a,b) for the s and d-wave momentum spectrum respectively. We find that at momenta around $1GeV/c$ for a deuteron with a longitudinal momentum of $1GeV/c$ -- a typical momentum anticipated for the deuteron beam of COSY --, the spectrum both in the s and d-wave is enhanced by typically one order of magnitude. In addition, the spherical symmetry in the s-state is increasingly broken with increasing $P_z$. This is most readily seen in coordinate space, where one obtains upon the corresponding Fourier transform
\begin{figure}[th]
\begin{center}
\includegraphics[scale=0.4,angle=-90]{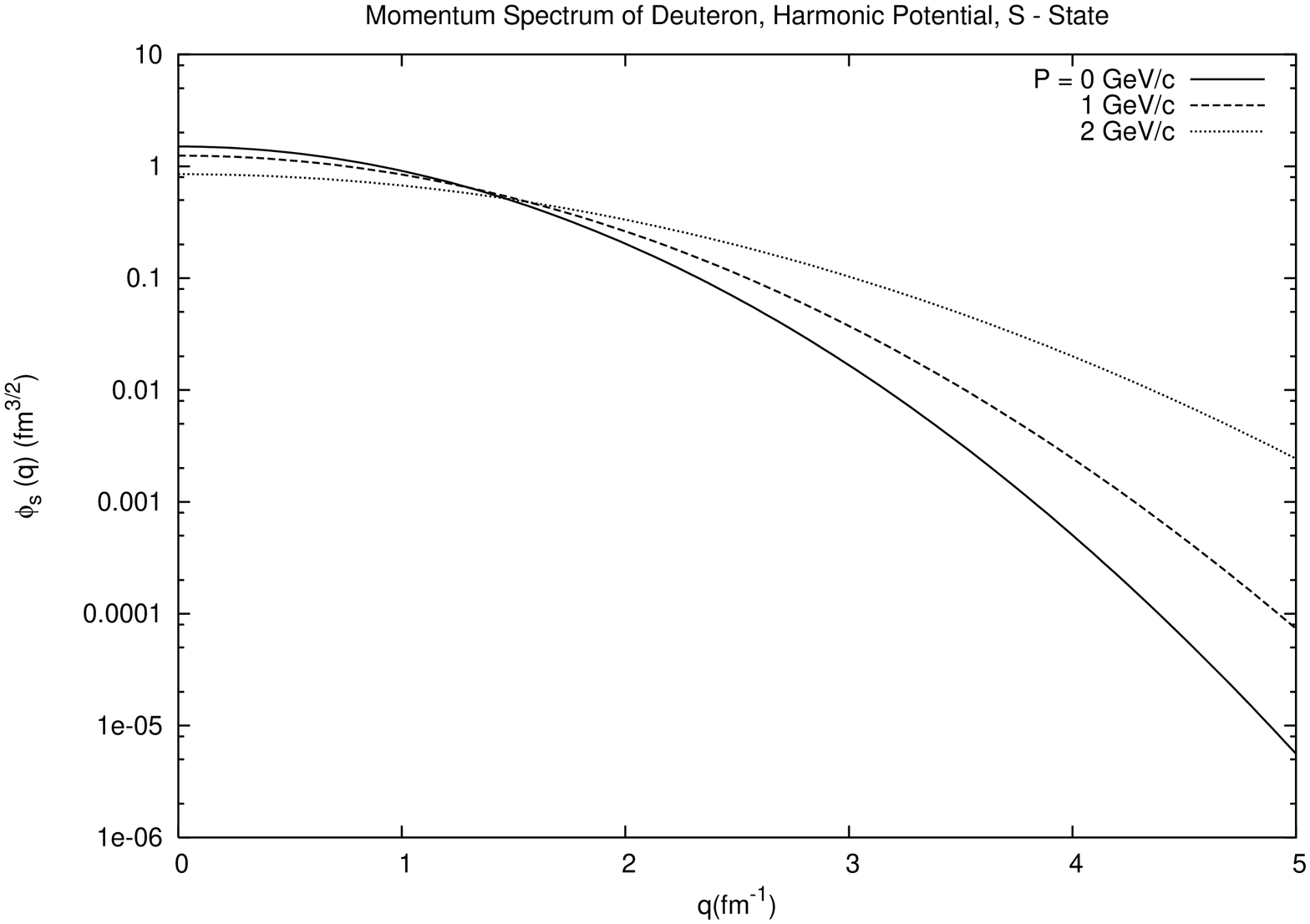}
\end{center}
\noindent{\footnotesize Fig. 1a. Momentum spectrum of the s-wave of the deuteron for different
external momenta $P_z$.}
\end{figure}
\begin{figure}[th]
\begin{center}
\includegraphics[scale=0.4,angle=-90]{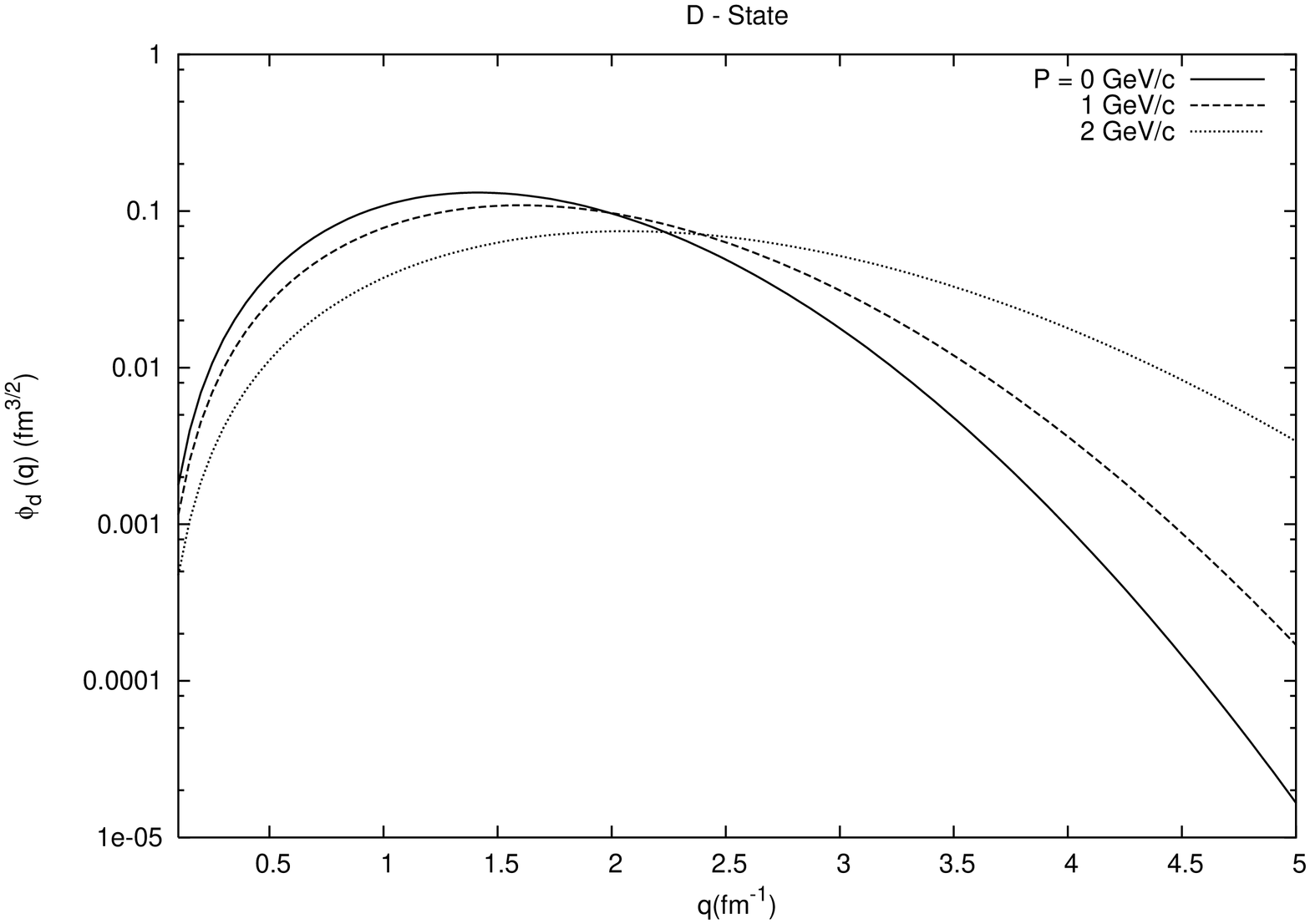}
\end{center}
\noindent{\footnotesize Fig. 1b. As Fig. 1a, however for the d-wave of the deuteron.}
\end{figure}
\begin{equation}
\phi_s(z,r)\sim e^{\frac{1}{2a^2}\left(\frac{M^2_d+P^2_z}{M^2_d}z^2+r_{\perp}^2\right)}
\end{equation}
or, if we introduce explicitly angular momentum functions,
\begin{equation}
\phi_s\sim e^{\frac{r^2}{2a^2}\left(1+\sqrt{\frac{16\pi}{45}}(Y_{20}(\hat{r})+
\frac{\sqrt{5}}{2}Y_{00}(\hat{r}))\frac{P_z^2}{M_d^2}\right)}\, .
\end{equation}
Clearly, as easily seen from an expansion of the exponential function, for $P_z>0$ the s-state wave
function for a deuteron contains in principle an infinite sum of even angular momentum components;
spin-orbital coupling admixes an additional L=2 component for the s-state wave function. A similar
qualitative argument holds also for the d-state (quantitatively, it differs by the additional
coupling of $Y_{2m}(\hat{r})$ from the structure of the d-state in the static deuteron).
\begin{figure}[th]
\begin{center}
\includegraphics[scale=0.4,angle=-90]{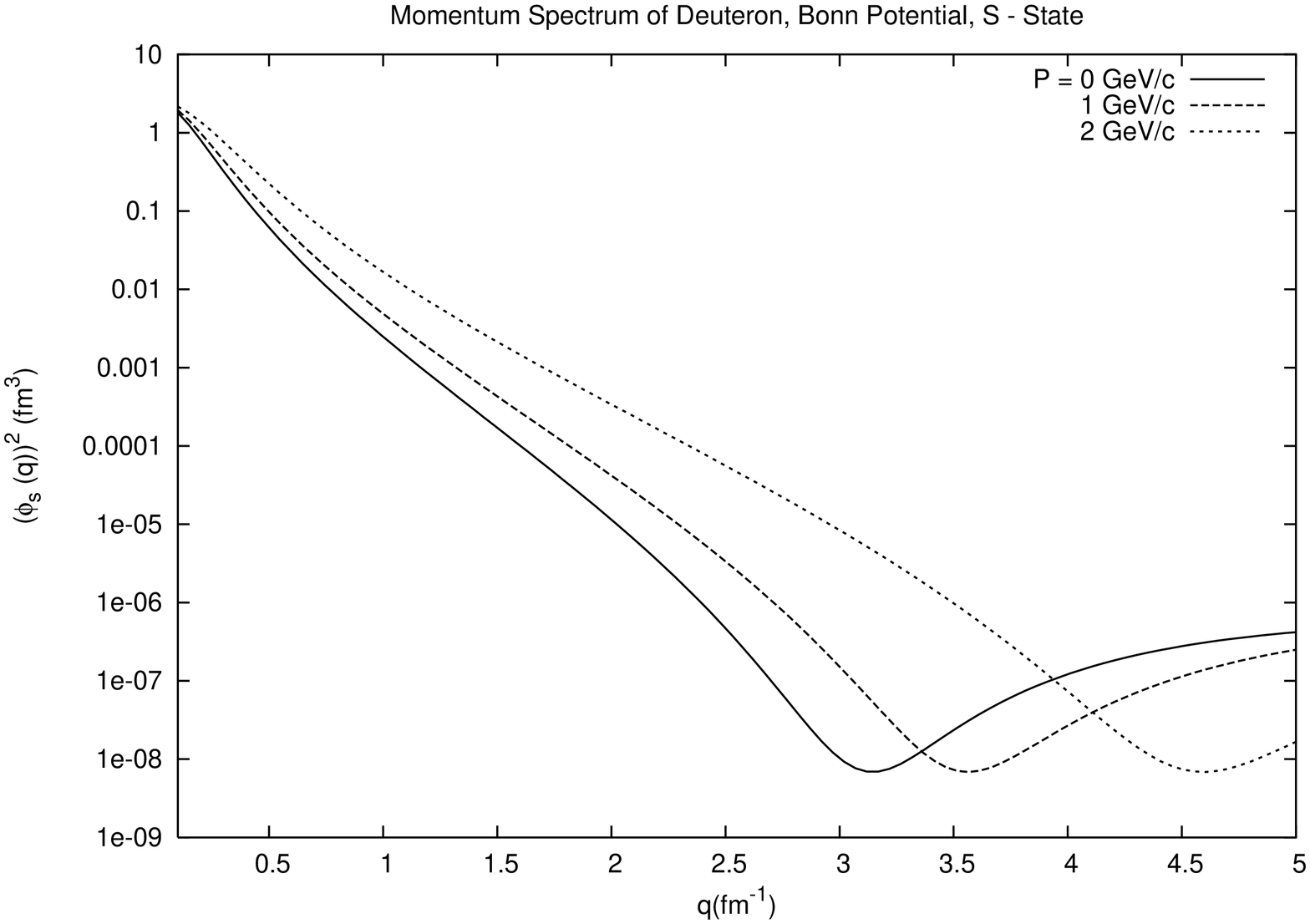}
\end{center}
\noindent{\footnotesize Fig. 2a. Momentum spectrum of the s-wave of the deuteron for different
external momenta $P_z$.}
\end{figure}
\begin{figure}[th]
\begin{center}
\includegraphics[scale=0.4,angle=-90]{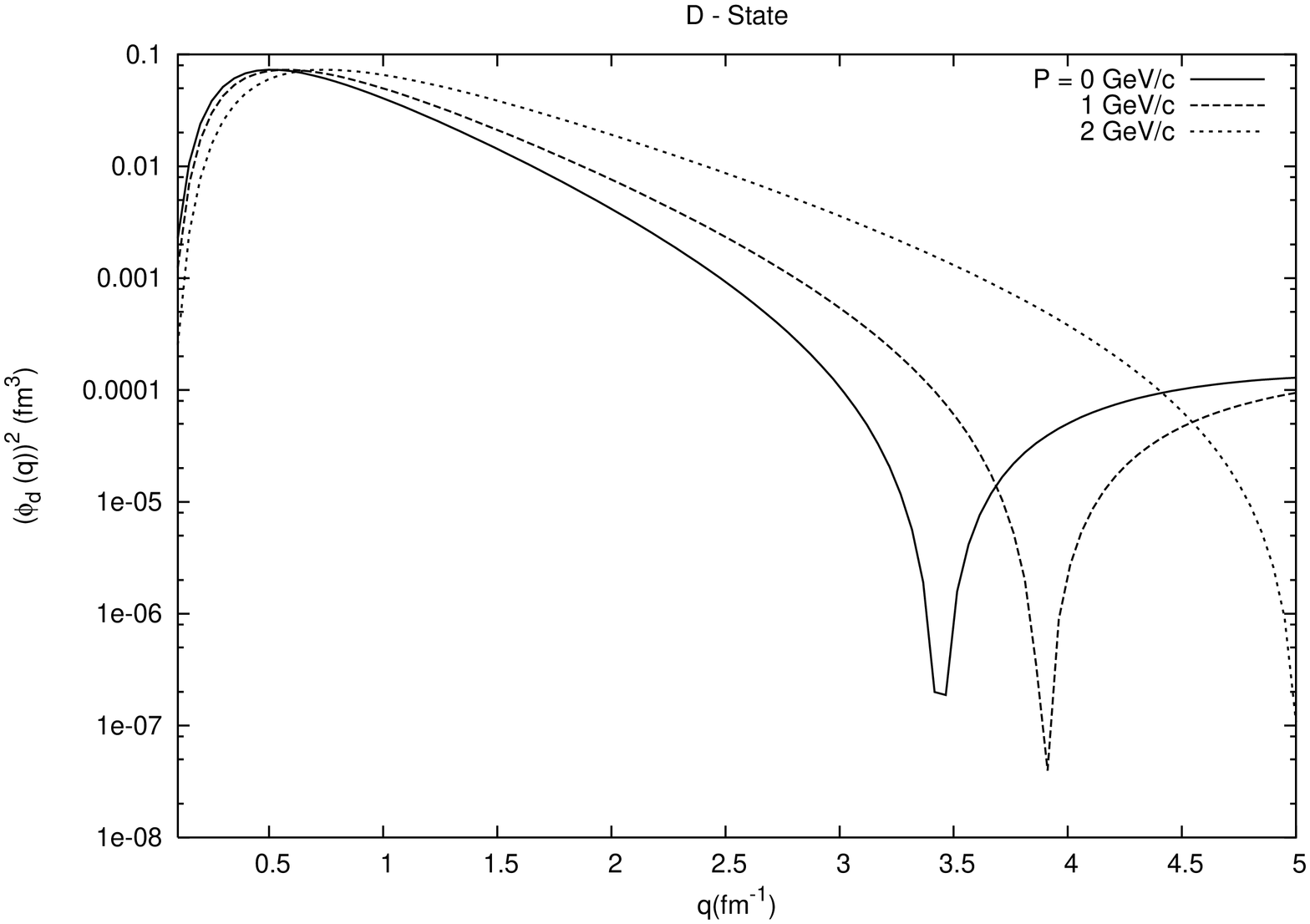}
\end{center}
\noindent{\footnotesize Fig. 2b. As Fig. 2a, however for the d-wave of the deuteron.}
\end{figure}
We extent our results for an harmonic kernel to deuteron wave functions from modern NN potentials, using the recipe $q^2_z \rightarrow \lambda^2(P_z) q^2_z$ explicitly for the deuteron wave function from the Bonn potential. The result is shown in Fig. 2(a,b) for the s and the d-state of the deuteron. Qualitatively, the resulting pattern - the enhancement of the $q_z$ spectrum for $P_z > 0$ - is similar as for an harmonic kernel, though quantitatively such a modification is less pronounced, as already for a deuteron in rest the Bonn potential significantly enhances the high momentum spectrum of the deuteron wave function relative to an harmonic potential.\\

\section{Summary, conclusions and outlook}
In this note we investigate the momentum spectrum of a deuteron in flight. Starting from the BSE and following the procedure of Blankenbecler and Sugar, we derive a 3-dimensional QPE for the deuteron. Introducing an harmonic kernel for the binding of the deuteron, we solve the corresponding equation for the s and d-state of the deuteron analytically.\\

As the main result we find, that for the longitudinal component (along the external momentum $P_z$) the momentum spectrum, if compared with a deuteron in rest or with the perpendicular components, is substantially modified with increasing $P_z$: the high momentum components are significantly enhanced, as expected from the quenching of the deuteron in coordinate space along the direction of motion. This is not surprising, as even in the s-state the spherical symmetry is broken for $P>0$ and higher partial waves $l>0$ modify the s-state (with $l=0$ for a deuteron in rest), whereas the conventional d-state develops in addition an l=0 component. Thus the deuteron gets more and more deformed with increasing $P_z$.\\

From the quantitative point of view our results are certainly preliminary: clearly our approach has to be based on a more realistic interaction kernel for comparison with a deuteron in rest, by taking the deuteron wave function or deuteron form factors from modern nucleon-nucleon potentials\cite{2Garcon}. A simple, but still approximate description is to fit for example the elastic deuteron form factor for a deuteron in rest by a set of Gaussians with different size parameters along the results in chapter 3 (the result is then still analytical). A more rigorous step would be to solve d-quenching for a kernel from modern nucleon-nucleon potentials, at the expense however, to end up with a complicated system of coupled differential equations due to the coupling of longitudinal and perpendicular components for $P_z> 0$, which can be solved only numerically, so that the transparency of the analytical solution is lost. In addition it has to be stressed that Lorentz boosts in the instant form are dynamical and involve the underlying interaction\cite{16Keister} (opposite to kinematical boots on the light cone \cite{17Brodsky}); thus there
exist different prescriptions for their incorporation \cite{18Plessas},\cite{19Des}. Experimentally, detailed information on Lorentz boosts of the deuteron will result a systematic study of pd and, in particular, dp reactions at COSY, where a (polarized) deuteron beam will be available in the very near future\cite{20COSY}.

\end{document}